\documentclass[aps,twocolumn,floatfix,showpacs,eqsecnum]{revtex4} 

\usepackage{graphicx}
\usepackage{amsmath}
\usepackage{amsfonts} 
\usepackage{bm}

\begin{document}

\title{Effects of disorder on the transmission of nodal fermions through a \textit{d}-wave superconductor}
\author{J. K. Asb\'{o}th$ ^{1,2}$, A. R. Akhmerov$ ^1$, M. V. Medvedyeva$ ^1$, and C. W. J. Beenakker$ ^1$}
\affiliation{$\phantom{1}^1$:Instituut-Lorentz, Universiteit Leiden, P.O. Box 9506, 2300 RA Leiden, The Netherlands\\
$\phantom{1}^2$:Research Institute for Solid State Physics and Optics, Hungarian Academy of Sciences, P.O. Box 49, H-1525 Budapest, Hungary}
\date{March 2011}
\begin{abstract}
The bulk microwave conductivity of a dirty \textit{d}-wave superconductor is known to depend sensitively on the range of the disorder potential: long-range scattering enhances the conductivity, while short-range scattering has no effect. Here we show that the three-terminal electrical conductance of a normal-metal--\textit{d}-wave superconductor--normal-metal junction has a dual behavior: short-range scattering suppresses the conductance, while long-range scattering has no effect.
\end{abstract}
\maketitle

\section{Introduction}
\label{intro}

As pointed out by Lee in an influential paper \cite{Lee93}, disorder has two competing effects on the microwave conductivity of a layered superconductor with \textit{d}-wave symmetry of the pair potential. On the one hand, disorder increases the density of low-energy quasiparticle excitations, located in the Brillouin zone near the intersection of the Fermi surface with the nodal lines of vanishing excitation gap. On the other hand, disorder reduces the mobility of these nodal fermions. For short-range scattering the two effects cancel \cite{Fra86}, producing a disorder independent microwave conductivity $\sigma_{0}\simeq(e^{2}/h)k_{F}\xi_{0}$ per layer in the low-temperature, low-frequency limit (with $\xi_{0}$ the coherence length and $k_{F}$ the Fermi wave vector). For long-range scattering the first of the two effects wins \cite{Dur00,Nun05}, which explains the conductivity enhancement measured in the high-$T_{c}$ cuprates \cite{Lee96,Hos99} (where long-range scattering dominates \cite{Dol98}).

The microwave conductivity is a bulk property of an unbounded system, of length $L$ and width $W$ large compared to the mean free path $l$. A finite system makes it possible to study the crossover from diffusive to ballistic transport, as $L$ and $W$ become smaller than $l$. We have recently shown \cite{Asb09} that the transmission of nodal fermions over a length $L$ in the range $\xi_{0}\ll L\ll l,W$ is pseudodiffusive: The transmission probability has the $W/L$ scaling of a diffusive system, even in the absence of any disorder. The corresponding conductance $G_{0}$ is close the value $(W/L)\sigma_{0}$ which one would expect from the microwave conductivity, up to a small correction of order $(k_{F}\xi_{0})^{-2}\ll 1$.

It is the purpose of this paper to investigate the effects of disorder on the pseudodiffusive conductance, as $L$ becomes larger than $l$. We find a qualitatively different behavior than for the microwave conductivity, with an exponentially suppressed conductance in the case of short-range scattering and an unaffected conductance $G\simeq G_{0}$ for long-range scattering.

\section{Formulation of the problem}
\label{formulation}

\begin{figure}[tb]
\centerline{\includegraphics[width=0.5\linewidth]{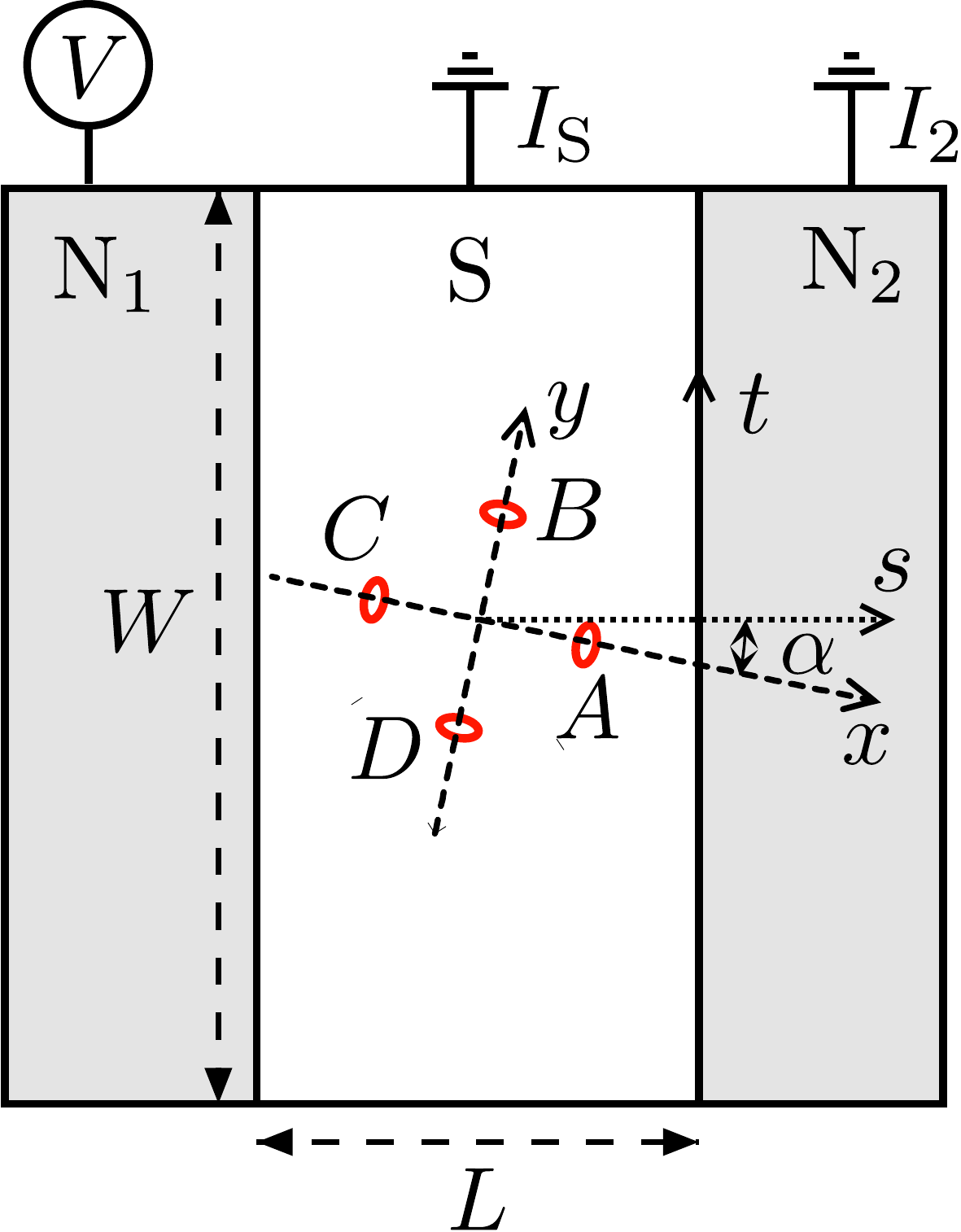}}
\caption{\label{fig_layout}
Geometry to measure the transmission of quasiparticles at the nodes (red circles) of the pair potential with $d_{xy}$ symmetry.
}
\end{figure}

The geometry to measure the transmission of nodal fermions is illustrated in Fig.\ \ref{fig_layout}. It consists of a superconducting strip S between two normal metal contacts ${\rm N}_{1}$ and ${\rm N}_{2}$. The transverse width $W$ of the superconductor is assumed to be large compare to the separation $L$ of the NS interfaces, in order to avoid edge effects. Contact ${\rm N}_{1}$ is at an elevated voltage $V$, while S and ${\rm N}_{2}$ are both grounded. The current $I_{2}$ through contact ${\rm N}_{2}$ measures the transmitted charge, which is carried entirely by nodal fermions if $L\gg\xi_{0}$. The nodal lines are the $x$ and $y$ axes, oriented at an angle $\alpha$ relative to the normal to the NS interfaces. There are four nodal points $A,B,C,D$ in the Brillouin zone, at the intersection of the nodal lines and the Fermi surface. The nodal fermions have an anisotropic dispersion relation, with a velocity $v_{F}$ parallel to the nodal axis and a much smaller velocity $v_{\Delta}=v_{F}/k_{F}\xi_{0}$ perpendicular to the nodal axis. 

The (three-terminal) conductance $G=I_{2}/V$ was calculated in Ref.\ \onlinecite{Asb09} in the clean limit $L\ll l$, with the result (per layer)
\begin{equation}
G_{\rm clean}=\frac{2e^{2}}{h}\frac{W}{L}\frac{v_{F}^{2}+v_{\Delta}^{2}}{\pi v_{F}v_{\Delta}}\frac{\Gamma_{1}\Gamma_{2}}{(2-\Gamma_{1})(2-\Gamma_{2})},\label{Gclean}
\end{equation}
independent of $\alpha$. The factors $\Gamma_{1},\Gamma_{2}\in(0,1)$ are the (mode-independent) transmission probabilities of tunnel barriers at the ${\rm N}_{1}{\rm S}$ and ${\rm N}_{2}{\rm S}$ interfaces. We have assumed that the tunnel barriers do not couple the nodes, which requires $\alpha\gg\xi_{0}/L$ and $\pi/4-\alpha\gg\xi_{0}/L$. Since $\xi_{0}/L\ll 1$, this is the generic case. 

We now wish to move away from the clean limit and include scattering by electrostatic potential fluctuations. We distinguish two regimes, depending on the magnitude of the correlation length $l_{c}$ of the potential fluctuations. In the regime $k_{F}l_{c}\gg 1$ of long-range disorder, the nodes remain uncoupled and can be treated separately. We consider this regime of \textit{intranode} scattering first, and then include the effects of \textit{internode} scattering when $l_{c}$ becomes smaller than $1/k_{F}$.

\section{Intranode scattering regime}
\label{intranoderegime} 

In the absence of internode scattering, the electron and hole components of the wave function $\Psi=(\Psi_{e},\Psi_{h})$ of nodal fermions (at excitation energy $\varepsilon$) are governed by the anisotropic Dirac equation $H\Psi=\varepsilon\Psi$. Near node $A$ the Hamiltonian takes the form \cite{Alt02}
\begin{equation}
H=-i\hbar(v_{F}\sigma_{z}\partial_x+v_{\Delta}\sigma_{x}\partial_y)+V_{\mu}\sigma_{z}+V_{\Delta}\sigma_{x}.\label{DiracA}
\end{equation}
The two terms $V_{\mu}(x,y)$ and $V_{\Delta}(x,y)$ describe, respectively, long-range disorder in the electrostatic potential and in the $s$-wave component of the pair potential. These two types of disorder preserve time-reversal symmetry. The Hamiltonian anti-commutes with the Pauli matrix $\sigma_{y}$, belonging to the chiral symmetry class AIII of Ref.\ \onlinecite{Alt02}.

Following Refs.\ \onlinecite{Sch09,Tit09}, at zero energy, the disorder potentials can be transformed out from the Dirac equation by means of the transformation $\Psi\mapsto\exp(i\phi+\chi\sigma_{y})\Psi_{0}$, with fields $\phi$ and $\chi$ determined by
\begin{subequations}
\label{phichidef}
\begin{align}
&v_{F}\partial_{x}\phi+v_{\Delta}\partial_{y}\chi=-V_{\mu}/\hbar,\label{phichi1}\\
&v_{F}\partial_{x}\chi-v_{\Delta}\partial_{y}\phi=V_{\Delta}/\hbar.\label{phichi2}
\end{align}
\end{subequations}
If $H\Psi=0$ then also $H_{0}\Psi_0=0$, where $H_{0}$ is the Dirac Hamiltonian without disorder 
($V_{\mu}\equiv 0$ and $V_{\Delta}\equiv 0$).

The transformation from $\Psi$ to $\Psi_{0}$ leaves the particle current density unaffected but not the electrical current density: The particle current density $\mathbf{j}$ reads
\begin{equation}
(j_x, j_y) = \Psi^{\dagger}(v_F \sigma_{z}, v_\Delta \sigma_x) \Psi=\Psi_{0}^{\dagger}(v_F \sigma_{z}, v_\Delta \sigma_x)\Psi_{0},\label{jparticledef}
\end{equation}
while for the electrical current density $\mathbf{i}$ one has
\begin{equation}
i_y=0,\;\; i_x = e v_F\Psi^{\dagger}\Psi =ev_{F}\Psi_{0}^{\dagger}\exp(2\chi\sigma_{y})\Psi_{0}.\label{ielectricaldef}
\end{equation}
This is consistent with the findings of Durst and Lee \cite{Dur00}, that the low-energy effects of intranode scattering on the density-of-states and on the mobility cancel for the thermal conductivity (proportional to the particle current) but not for the electrical conductivity (which is increased by disorder).

As we now show, for the conductance of a finite system, the effect of intranode scattering is entirely different. Following Ref.\ \cite{Asb09}, the conductance is determined by the transfer matrix ${\cal M}$ relating right-moving and left-moving states $\Phi_{1}=(\Phi^{+}_{1},\Phi^{-}_{1})$ in ${\rm N}_{1}$ to right-moving and left-moving states $\Phi_{2}=(\Phi^{+}_{2},\Phi^{-}_{2})$ in ${\rm N}_{2}$. It is convenient to rotate the coordinate system from $x$ and $y$ along the nodal axes to coordinates $s$ and $t$ perpendicular and parallel to the NS interfaces. The transfer matrix is defined by 
\begin{equation}
\Phi_{2}(L,t)=\int dt'\,{\cal M}(t,t')\Phi_{1}(0,t').\label{PsiMrelation}
\end{equation}
For wave vectors in the normal metal coupled to node $A$, the right-movers are electrons $\Phi_{e}^{+}$ and the left-movers are holes $\Phi_{h}^{-}$, so an electron incident from contact ${\rm N}_{1}$ can only be transmitted into contact ${\rm N}_{2}$ as an electron, not as a hole. The corresponding transmission matrix $t_{ee}$ is determined by the transfer matrix via
\begin{equation}
t_{ee}=\left({\cal M}_{11}^{\dagger}\right)^{-1},\;\;
{\cal M}=
\begin{pmatrix}
{\cal M}_{11}&{\cal M}_{12}\\
{\cal M}_{21}&{\cal M}_{22}
\end{pmatrix}.
\label{teeM11}
\end{equation}
The contribution $G_{A}$ to the electrical conductance from node $A$ then follows from
\begin{equation}
G_{A}=\frac{2e^{2}}{h}{\rm Tr}\,t_{ee}^{\vphantom{\dagger}}t_{ee}^{\dagger},\label{GA}
\end{equation}
with a factor of two to account for both spin directions. The full conductance contains an additional contribution from node $B$, determined by similar expressions with $\alpha$ replaced by $\alpha-\pi/2$.

The Hamiltonian \eqref{DiracA} does not apply within a coherence length $\xi_{0}$ from the NS interfaces, where the depletion of the pair potential should be taken into account. We assume weak disorder, $l\gg\xi_{0}$, so that we can use the clean-limit results of Ref.\ \cite{Asb09} in this interface region. For simplicity, we do not include tunnel barriers at this stage ($\Gamma_{1}=\Gamma_{2}=1$). The transfer matrix through the superconductor is then given by
\begin{align}
{\cal M}={}&\exp(i\phi_{\rm R}+\sigma_{y}\chi_{\rm R})\exp(-iLv_{F}v_{\Delta}v_{\alpha}^{-2}\sigma_{y}\partial_{t}+L\varphi_{\alpha}\partial_{t})\nonumber\\
&\times\exp(-i\phi_{\rm L}-\sigma_{y}\chi_{\rm L}),\label{Mresult}
\end{align}
with the abbreviations
\begin{align}
&v_{\alpha}=\sqrt{v_{F}^{2}\cos^{2}\alpha+v_{\Delta}^{2}\sin^{2}\alpha},\label{valphadef}\\
&\varphi_{\alpha}=\tfrac{1}{2}v_{\alpha}^{-2}(v_{F}^{2}-v_{\Delta}^{2})\sin 2\alpha.\label{phialpahdef}
\end{align}
The fields $\phi_{\rm L}(t),\chi_{\rm L}(t)$ are evaluated at the left NS interface ($s=0$) and the fields $\phi_{\rm R}(t),\chi_{\rm R}(t)$ are evaluated at the right NS interface ($s=L$).

We now follow Ref.\ \cite{Sch09} and use the freedom to impose boundary conditions on the solution of Eq.\ \eqref{phichidef}. Demanding $\chi=0$ on the NS interfaces fixes both $\chi$ and $\phi$ (up to an additive constant). The transfer matrix \eqref{Mresult} then only depends on the disorder through the terms $\exp(i\phi_{\rm R})$ and $\exp(-i\phi_{\rm L})$, which are unitary transformations and therefore drop out of the conductance \eqref{GA}. We conclude that the electrical conductance \eqref{Gclean} is not affected by long-range disorder.

Tunnel barriers affect the conductance in two distinct ways. Firstly, at both NS interfaces, we need to consider all four states $\Phi_{e,h}^{\pm}$ that have the same component of the wave vector parallel to the NS interface ($\Phi_{e}^{+},\Phi_{h}^{-}$ have the opposite perpendicular component than $\Phi_{e}^{-},\Phi_{h}^{+}$). However, only one right-moving and one left-moving superposition of these modes, $\Phi_{n}^{\pm}$, is coupled by the transfer matrix to the other side of the system: 
\begin{subequations}
\label{Phieh}
\begin{align}
&\Phi_{n}^{+}=(2-\Gamma_{n})^{-1/2}\bigl[\Phi_{e}^{+}+(1-\Gamma_{n})^{1/2}\Phi_{h}^{+}\bigr],\label{Phieha}\\
&\Phi_{n}^{-}=(2-\Gamma_{n})^{-1/2}\bigl[(1-\Gamma_{n})^{1/2}\Phi_{e}^{-}+\Phi_{h}^{-}\bigr].\label{Phiehb}
\end{align}
\end{subequations}
The superposition of incoming electron and hole states orthogonal to $\Phi_n^{+}$ is fully reflected by the tunnel barrier and the superconductor, and so plays no role in the conductance.
For a detailed derivation of these formulas see Appendix \ref{sect:tunnel}. 

Secondly, the modes $\Phi_{n}^{+}$ are only partially transmitted through the barriers. We have calculated the transmission probability (see Appendix \ref{sect:tunnel} for details), and found that it can be accounted for by the following transformation of the transfer matrix,
\begin{equation}
{\cal M}\mapsto e^{\gamma_{2}\sigma_{y}}{\cal M}e^{\gamma_{1}\sigma_{y}},\;\;\gamma_{n}=\tfrac{1}{2}\ln\bigl(2/\Gamma_{n}-1\bigr).\label{Mgamma}
\end{equation}

With tunnel barriers, the transmission matrix contains mixed electron and hole elements,
\begin{equation}
{\cal T}=\begin{pmatrix}
t_{ee}&t_{eh}\\
t_{he}&t_{hh}
\end{pmatrix}=U_{2}^{\dagger}\begin{pmatrix}
({\cal M}_{11}^{\dagger})^{-1}&0\\
0&0
\end{pmatrix}U_{1},\label{tUMrelation}
\end{equation}
where the unitary matrices $U_{n}$ transform from the electron-hole basis to the basis state $\Phi_{n}^{+}$ and its (fully reflected) orthogonal complement,
\begin{equation}
U_{n}=(2-\Gamma_{n})^{-1/2}\begin{pmatrix}
1&(1-\Gamma_{n})^{1/2}\\
(1-\Gamma_{n})^{1/2}&-1
\end{pmatrix}.\label{Undef}
\end{equation}
Finally, the contribution $G_{A}$ to the electrical conductance from node $A$ follows from
\begin{equation}
G_{A}=\frac{2e^{2}}{h}{\rm Tr}\,\bigl(t_{ee}^{\vphantom{\dagger}}t_{ee}^{\dagger}-t_{he}^{\vphantom{\dagger}}t_{he}^{\dagger}\bigr).\label{GAtunnel}
\end{equation}
With tunnel barriers, not just nodes $A$ and $B$, but nodes $C$ and $D$ also contribute to the full conductance.

Collecting results, we substitute Eq.\ \eqref{Mresult} (with $\chi_{\rm L}$ and $\chi_{\rm R}$ both fixed at zero) into Eq.\ \eqref{Mgamma} to obtain the transfer matrix, and then substitute the $1,1$ block into Eq.\ \eqref{tUMrelation} for the transmission matrix. Disorder only enters through the factors $\exp(i\phi_{\rm R})$ and $\exp(-i\phi_{\rm L})$, which mix the modes on the superconducting side of the tunnel barriers. Since the tunnel probabilities are assumed to be mode independent, these factors commute with the $U_{n}$'s and cancel upon taking the trace in Eq.\ \eqref{GAtunnel}. We thus recover the clean-limit result \eqref{Gclean}, independent of any disorder potential. Disorder would have an effect on the conductance for mode-dependent tunnel probabilities, but since the modes in the normal metal couple to a narrow range of transverse wave vectors in the superconductor, the assumption of mode-independence is well justified. 

As an aside we mention that the thermal (rather than electrical) conductance $G_{\rm thermal}\propto{\rm Tr}\,{\cal T}{\cal T}^{\dagger}$ would be independent of disorder also for the case of mode-dependent tunnel probabilities, since the $U_{n}$'s drop out of the trace. The tunnel barriers would then still enter in the transfer matrix through the terms $e^{\gamma_{n}\sigma_{y}}$ in Eq.\ \eqref{Mgamma}, but these terms have the same effect as delta function contributions to $V_{\mu}$ and can therefore be removed by including them in Eq.\ \eqref{phichidef}. The conclusion is that the thermal conductance is independent of both disorder and tunnel barriers, while the electrical conductance is independent of disorder but dependent on tunnel barriers through the factors $\Gamma_{n}/(2-\Gamma_{n})$. Notice that the Wiedemann-Franz relation between thermal and electrical conductance does not apply.

\section{Effect of internode scattering}
\label{internoderegime} 

So far we have only considered intranode scattering. For short-range disorder we have to include also the effects of internode scattering. Internode scattering suppresses the electrical conductance, measured between the normal metals ${\rm N}_{1}$ and ${\rm N}_{2}$, because an electron injected from ${\rm N}_{1}$ into nodes $A$ or $B$ and then scattered to nodes $C$ or $D$ will exit into ${\rm N}_{2}$ as a hole, of opposite electrical charge. (The charge deficit is drained to ground via the superconductor.) The thermal conductance, in contrast, remains unaffected by internode scattering because electrons and holes transport the same amount of energy. (Again, the Wiedemann-Franz relation does not apply.)

We first give a semiclassical analytical theory, and then a fully quantum mechanical numerical treatment.

\subsection{Semiclassical theory}
\label{semiclassics}

We assume that the mean free path $l$ for intranode scattering is short compared to the internode scattering length. Semiclassically we may then describe the internode scattering by a (stationary) reaction-diffusion equation for the carrier densities $n_{\nu}$,
\begin{equation}
\bm{\nabla}\cdot{\bm D}_{\nu}\cdot\bm{\nabla}n_{\nu}+\sum_{\nu'\neq \nu}\bigl(\gamma_{\nu\nu'}n_{\nu'}-\gamma_{\nu'\nu}n_{\nu}\bigr)=0.\label{reactiondiff}
\end{equation}
The labels $\nu,\nu'\in\{A,B,C,D\}$ indicate the nodes, with diffusion tensor ${\bm D}_{\nu}$ and scattering rate $\gamma_{\nu\nu'}$ from $\nu'$ to $\nu$. For simplicity we assume there is no tunnel barrier at the NS interfaces, and seek a solution $n_{\nu}(s)$ with boundary conditions
\begin{equation}
n_{\nu}(0)=\frac{1}{2}(\delta_{\nu,A}+\delta_{\nu,B})eV\rho_{F},\;\;n_{\nu}(L)=0. \label{boundarycond}
\end{equation}
Here $\rho_{F}$ is the density of states per node at the Fermi energy, and we have chosen the sign of the applied voltage $V$ such that electrons (rather than holes) are injected into the superconductor from ${\rm N}_{1}$. 

The diffusion tensor is diagonal in the $x-y$ basis, with components $D_{\mu}$ and $D_{\Delta}$ in the direction of $v_{\mu}$ and $v_{\Delta}$, respectively. The average diffusion constant is $\bar{D}=\frac{1}{2}(D_{\mu}+D_{\Delta})$ and we also define $D_{\alpha}=D_{\mu}\cos^{2}\alpha+D_{\Delta}\sin^{2}\alpha$. We distinguish internode scattering between opposite nodes, with rate $\gamma_{1}$, and between adjacent nodes, with rate $\gamma_{2}$. Because the solution $n_{\nu}(s)$ in the $s-t$ basis is independent of the transverse coordinate $t$, we may replace the Laplacian $\bm{\nabla}\cdot{\bm D}_{\nu}\cdot\bm{\nabla}\mapsto D_{\nu}d^{2}/ds^{2}$ with $D_{A}=D_{C}=D_{\alpha}$ and $D_{B}=D_{D}=2\bar{D}-D_{\alpha}$. 

We seek the current into ${\rm N}_{2}$, given by
\begin{equation}
I_{2}=-eW\lim_{s\rightarrow L}\frac{d}{ds}\bigl[D_{A}n_{A}+D_{B}n_{B}-D_{C}n_{C}-D_{D}n_{D}\bigr].\label{I2def}
\end{equation}
This can be obtained by integrating the reaction-diffusion equation \eqref{reactiondiff} in the way explained in Ref.\ \onlinecite{Sta86}. The result is
\begin{align}
I_{2}={}&e^{2}V\rho_{F}W\frac{1}{2}\biggl[\frac{\sqrt{2(\gamma_{1}+\gamma_{2})D_{\alpha}}}{\sinh\sqrt{2L^{2}(\gamma_{1}+\gamma_{2})/D_{\alpha}}}\nonumber\\
&+\frac{\sqrt{2(\gamma_{1}+\gamma_{2})(2\bar{D}-D_{\alpha})}}{\sinh\sqrt{2L^{2}(\gamma_{1}+\gamma_{2})/(2\bar{D}-D_{\alpha})}}\biggr].\label{I_2result}
\end{align}
In the small-$L$ limit (when intervalley scattering can be neglected) we recover an $\alpha$-independent conductance $I_{2}/V\rightarrow e^{2}\rho_{F}\bar{D}W/L$, consistent with the expected result \eqref{Gclean}. For large $L$ the conductance decays exponentially $\propto e^{-L/l_{\rm inter}}$, with
\begin{equation}
l_{\rm inter}=\sqrt{\tfrac{1}{2}\min(D_{\alpha},2\bar{D}-D_{\alpha})/(\gamma_{1}+\gamma_{2})}\label{linterdef}
\end{equation}
the internode scattering length. For weak disorder ($k_{F}l\gg 1$) this decay length is much shorter than the Anderson localization length $\simeq le^{k_{F}l}$, so we are justified in treating the transport semiclassically by a diffusion equation.

\subsection{Fully quantum mechanical solution}
\label{numerics}

The Hamiltonian in the presence of internode scattering belongs to symmetry class CI of Ref.\ \cite{Alt02}, restricted by time-reversal symmetry and electron-hole symmetry --- but without the chiral symmetry that exists in the absence of internode scattering. 

To write the Hamiltonian ${\cal H}$ of the four coupled nodes in a compact form we use three sets of Pauli matrices: For each $i=x,y,z$ the $2\times 2$ Pauli matrix $\sigma_{i}$ couples electrons and holes, $\gamma_{i}$ couples opposite nodes ($A$ to $C$ and $B$ to $D$), and $\tau_{i}$ couples adjacent nodes ($A$ to $B$ and $C$ to $D$). The requirements of time-reversal symmetry and electron-hole symmetry are given, respectively, by
\begin{equation}
\gamma_{x}{\cal H}^{\ast}\gamma_{x}={\cal H},\;\;
(\gamma_{x}\otimes\sigma_{y}){\cal H}^{\ast}(\gamma_{x}\otimes\sigma_{y})=-{\cal H}.\label{CTsym}
\end{equation}

In the absence of disorder, the Hamiltonian is given by
\begin{align}
{\cal H}_{\rm clean}={}&p_{x} \left(v_{F}\tau_{+}\otimes\sigma_z +
v_\Delta\tau_{-}\otimes\sigma_x\right)\otimes\gamma_z\nonumber\\
&+p_{y}\left(v_{F}\tau_{-}\otimes\sigma_z +
v_\Delta\tau_{+}\otimes\sigma_x\right)\otimes\gamma_z.\label{H0def}
\end{align}
The momentum operator is $\bm{p}=-i\hbar\partial/\partial\bm{r}$ and we have defined $\tau_{\pm}=\frac{1}{2}(\tau_{0}\pm\tau_{z})$, with $\tau_{0}$ the $2\times 2$ unit matrix.

Since the effects of disorder in the electrostatic potential $V_{\mu}(\bm{r})$ and in the pair potential $V_{\Delta}(\bm{r})$ are equivalent \cite{Alt02}, we restrict ourselves to the former. The relevant Fourier components of $V_{\mu}(\bm{r})$ can be represented by the expansion
\begin{align}
V_{\mu}(\bm{r})=&{}\mu_{0}(\bm{r})\nonumber\\
&+\mu_{1}(\bm{r})e^{i(\bm{k}_{C}-\bm{k}_{A})\cdot\bm{r}}+\mu_{2}(\bm{r})e^{i(\bm{k}_{D}-\bm{k}_{B})\cdot\bm{r}}\nonumber\\
&+\mu_{3}(\bm{r})e^{i(\bm{k}_{B}-\bm{k}_{A})\cdot\bm{r}}+\mu_{4}(\bm{r})e^{i(\bm{k}_{C}-\bm{k}_{B})\cdot\bm{r}},\label{Vmuexpansion}
\end{align}
where $\bm{k}_{X}$ is the wave vector of node $X=A,B,C,D$ (see Fig.\ \ref{fig_layout}). The Fourier amplitudes $\mu_{p}(\bm{r})$ are all slowly varying functions of $\bm{r}$, with correlation length $\xi\gg 1/k_{F}$. The amplitude $\mu_{0}$ is responsible for intranode scattering, arising from spatial Fourier components of $V(\bm{r})$ with wave vector $\ll k_{F}$ (long-range scattering). The other four amplitudes arise from Fourier components with wave vector $\gtrsim k_{F}$ (short-range scattering).  Of these internode scattering potentials, $\mu_{1},\mu_{2}$ scatter between opposite nodes and $\mu_{3},\mu_{4}$ scatter between adjacent nodes.

The Hamiltonian ${\cal H}={\cal H}_{\rm clean}+{\cal H}_{\rm disorder}$ contains an electrostatic disorder contribution ${\cal H}_{\rm disorder}\propto\sigma_{z}$. Six combinations of Pauli matrices are allowed by the symmetry \eqref{CTsym}, five of which have independent amplitudes:
\begin{align}
&{\cal H}_{\rm disorder}=\sum_{n=0}^{4}{\cal H}_{p}\otimes\sigma_{z},\;\;{\rm with}\label{Hpdef}\\
{\cal H}_0&= \mu_{0}(\bm{r})\left[\tau_{+}\otimes\gamma_0+ \tau_{-}\otimes\gamma_0\right]=\mu_{0}(\bm{r})\tau_{0}\otimes\gamma_{0},\nonumber\\
{\cal H}_1&= \mu_{1}(\bm{r}) \tau_{+}\otimes\gamma_x,\;\;
{\cal H}_2= \mu_{2}(\bm{r}) \tau_{-}\otimes\gamma_x,\nonumber\\
{\cal H}_3&= \mu_{3}(\bm{r}) \tau_x\otimes\gamma_0,\;\;
{\cal H}_4= \mu_{4}(\bm{r}) \tau_x\otimes\gamma_x.
\end{align}

\begin{figure}
\includegraphics[width=\linewidth]{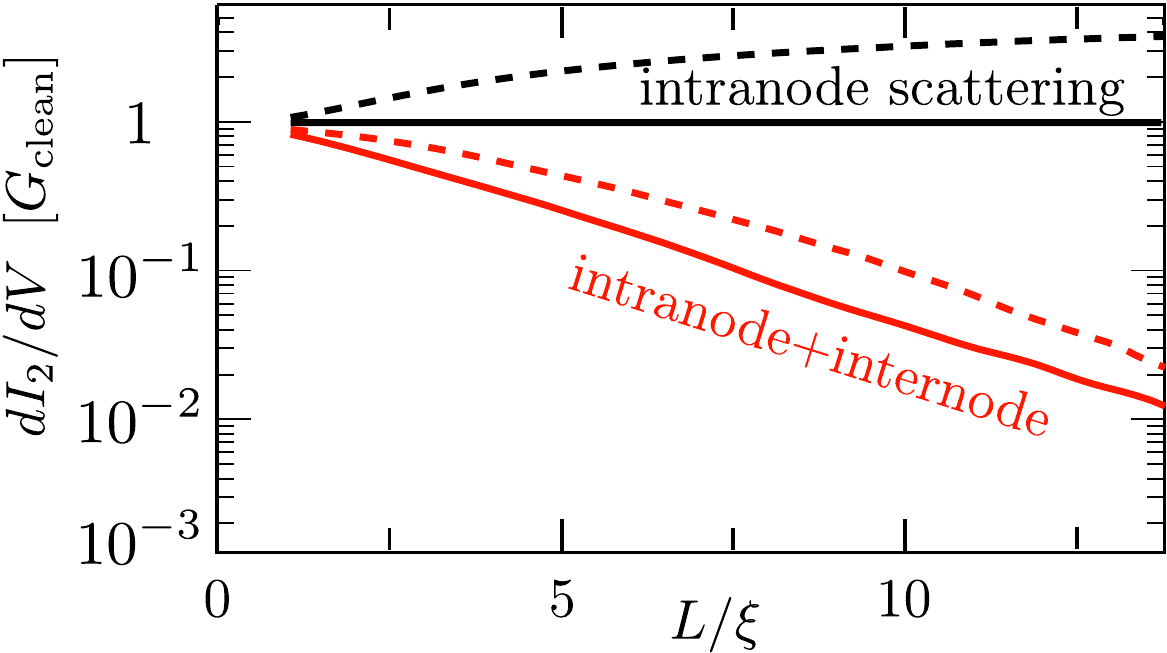}
\caption{
Differential conductance as a function of sample length, calculated numerically from the four coupled Dirac Hamiltonians of nodal fermions. The solid curves are at zero voltage and the dashed curves at nonzero voltage. If only intranode scattering is present (upper curves), the differential conductance is close to the value $G_{\rm clean}$ from Eq.\ \eqref{Gclean}. Including also internode scattering (lower curves) causes the conductance to decay strongly below $G_{\rm clean}$.
}
\label{fig:finite_e}
\end{figure}

We have solved the quantum mechanical scattering problem of the four coupled Dirac Hamiltonians numerically, by discretizing ${\cal H}$ on a grid. Since the electrostatic potential appears in the form of a vector potential in the Dirac Hamiltonian, in our numerical discretization we are faced with a notorious problem from the theory of lattice fermions: How to avoid fermion doubling while preserving gauge invariance \cite{Sta83}. The transfer matrix discretization method we use, from Ref.\ \cite{Jens}, satisfies gauge invariance only in the continuum limit. We ensure that we have reached that limit, by reducing the mesh size of the grid until the results have converged.

We fixed the width of the \textit{d}-wave strip at $W=150\,\xi$, oriented at an angle $\alpha=\pi/8$ with the nodal lines, and increased $L$ at fixed $\xi$. We set the anisotropy at $v_F/v_\Delta=2$ and did not include tunnel barriers for simplicity. All five amplitudes $\mu_{p}(\bm{r})$ are taken as independently fluctuating Gaussian fields, with the same correlation length $\xi$. The Gaussian fields have zero ensemble average, $\langle\mu_{p}(\bm{r})\rangle=0$, and second moment
\begin{equation}
K_{p}=(\hbar v_{F})^{-2}\int d\bm{r}\,\langle \mu_{p}(0)\mu_{p}(\bm{r})\rangle.\label{K0def}
\end{equation}
We took $K_{0}=1$ and either $K_{1}=K_{2}=K_{3}=K_{4}=0$ (only intranode scattering) or $K_{1}=K_{2}=K_{3}=K_{4}=0.4$ (both intranode and internode scattering). The results in Fig.\ \ref{fig:finite_e} give the differential conductance $dI_{2}/dV$, both at zero voltage and at a voltage of $V=0.2\,\hbar v_{F}/e\xi$.

Without internode scattering, we recover precisely the analytical result $dI_{2}/dV=G_{\rm clean}$ at $V=0$. At nonzero voltages, $dI_{2}/dV$ rises above $G_{\rm clean}$ with increasing $L$, consistent with the expectations \cite{Sch09} for the crossover from pseudo-diffusive to ballistic conduction at $V\simeq \hbar v_{F}/eL$. Internode scattering causes $dI_{2}/dV$ to drop strongly below $G_{\rm clean}$ with increasing $L$, both at zero and at nonzero voltages. The decay is approximately exponential, consistent with our semiclassical theory (although the range accessible numerically is not large enough to accurately extract a decay rate).

\section{Conclusion}
\label{conclude}

In summary, we have shown that the effect of disorder on the electrical current transmitted through a normal-metal--\textit{d}-wave-superconductor--normal metal junction is strikingly different depending on the range of the disorder potential: Long-range scattering has no effect, while short-range scattering suppresses the current exponentially. This behavior is dual to what is known \cite{Dur00,Nun05} for the bulk conductivity, which is unaffected by short-range scattering and increased by long-range scattering. Because of the exponential sensitivity $\propto e^{-L/l_{\rm inter}}$, we propose the setup of Fig.\ \ref{fig_layout} as a way to measure the internode scattering length $l_{\rm inter}$.

As a direction for future research, it would be interesting to study the transmission in the geometry of Fig.\ \ref{fig_layout} of low-energy excitations that are not located near the nodal points of the pair potential. A mechanism for the formation of non-nodal zero-energy states in \textit{d}-wave superconductors has been studied in Refs.\ \cite{Ada02,Ada04}.

\acknowledgments

We have benefited from discussions with I. Adagideli and J. Tworzyd{\l}o. This research was supported by the Dutch Science Foundation NWO/FOM and by an ERC Advanced Investigator Grant.

\appendix
\section{Tunnel barrier at the NS interface}
\label{sect:tunnel}
\begin{figure}
\includegraphics[width=\linewidth]{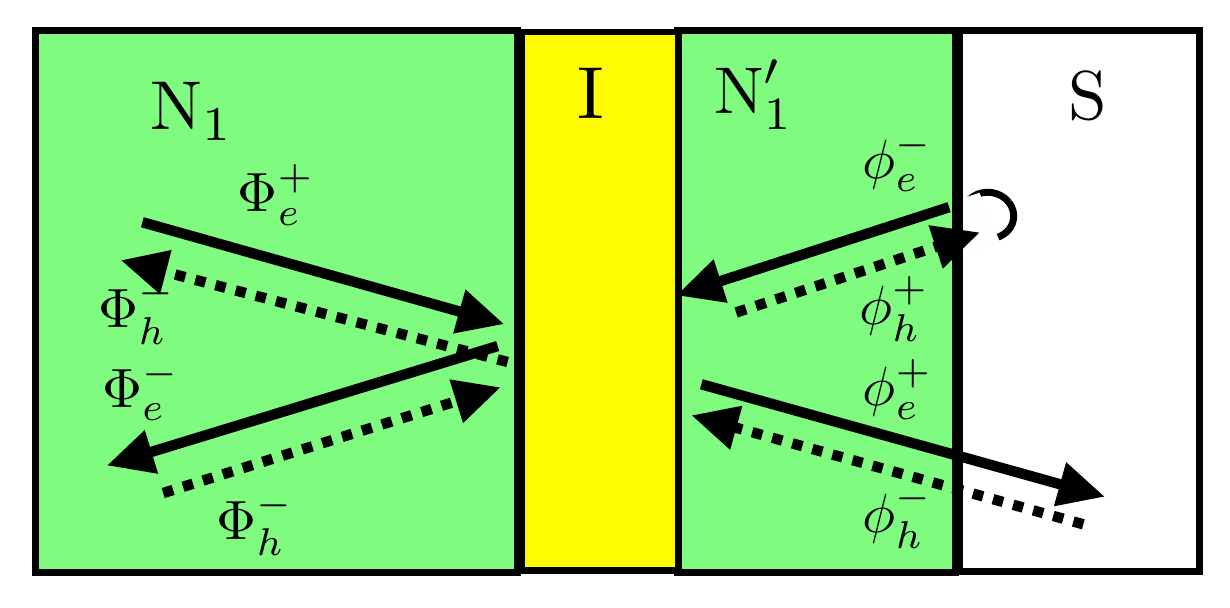}
\caption{Sketch of the normal-superconducting interface, with the
  plane wave modes taking part in conduction with a fixed energy and
  transverse momentum. To define the modes $\phi_{e,h}^{+,-}$, a piece of normal metal with length $\to 0$ is
  inserted between the tunnel barrier $\mathrm{I}$ and the
  superconductor $\mathrm{S}$.  
}
\label{fig:extra_n}
\end{figure}

We consider a tunnel barrier between the normal metal contact
$\mathrm{N}_1$ and the superconductor. To be specific, we describe the
left end of our setup, the derivations for the right contact follow
analogously. We introduce an additional
normal metal of zero length between the tunnel barrier and the
superconductor, as illustrated in Fig.~\ref{fig:extra_n}. 
For simplicity we assume translation invariance along the NS interface
holds: then the energy and the wave number along the NS interface are
good quantum numbers. The
tunnel barrier mixes the 4 modes 
with these constants in the normal lead
$\mathrm{N}_1$: $\Phi_e^{+,-}$ for right-/left-propagating electrons, and
$\Phi_h^{+-}$ for right-/left-propagating holes, with the 4 modes with
these constants in $\mathrm{N_1'}$: $\phi_e^{+,-}$ and
$\phi_h^{+,-}$. We have
\begin{align}
\begin{pmatrix} \Phi^-_e\\ \phi^+_e \\ \Phi^-_h\\ \phi^+_h \end{pmatrix}
&=
\begin{pmatrix} \mathfrak{r} & \mathfrak{t}' & 0 & 0 \\ \mathfrak{t} & \mathfrak{r}' & 0 & 0 \\
 0 & 0 & \mathfrak{r}^\ast & \mathfrak{t}'^\ast\\ 0 & 0 & \mathfrak{t}^\ast & \mathfrak{r}'^\ast \end{pmatrix}
\begin{pmatrix} \Phi^+_e\\ \phi^-_e \\ \Phi^+_h\\ \phi^-_h \end{pmatrix}.
\end{align}
Here $\mathfrak{t} = \sqrt{\Gamma_1}e^{i\chi}$ and $\mathfrak{t}'=\sqrt{\Gamma_1}
e^{i\chi'}$ are the electron transmission amplitudes, $\chi, \chi' \in
\mathbb{R}$, and $\mathfrak{r}$ and $\mathfrak{r}'$ are the electron reflection
amplitudes.

Since the angle $\alpha$ between the normal to the NS interface and the nodal line is
taken to be generic, $0\ll\alpha\ll-\pi/4$, the modes $\phi_h^+$ and
$\phi_e^-$ cannot propagate in the superconductor.  They are
localized near the NS interface, and follow Andreev reflection:
$\phi_e^- = -i \phi_h^+$. Using this, we can write the scattering
matrix $S$ representing the combined effect of the tunnel barrier and the
Andreev reflections on the propagating modes as
\begin{align} 
\begin{pmatrix}
\Phi_e^-\\ \Phi_h^- \\ \phi_e^+ 
\end{pmatrix}
&= S 
\begin{pmatrix}
\Phi_e^+\\ \Phi_h^+ \\ \phi_h^- 
\end{pmatrix}; 
\quad 
S = 
\begin{pmatrix}
\mathfrak{r}& - \mathfrak{t}' i {\mathfrak{t}'} &  - \mathfrak{t}' i {{\mathfrak{r}'}^\ast}\\ 
0       &\mathfrak{r}'                  &  {\mathfrak{t}'}^\ast \\
\mathfrak{t}& - \mathfrak{r}' i {\mathfrak{t}'}  &  -\mathfrak{r}' i {{\mathfrak{r}'}^\ast}
\end{pmatrix}
\label{eq:smatrix1}
\end{align} 

Now there are two incoming propagating modes from the left, but only
one outgoing propagating mode
to the right.  This implies that there is a superposition of
$\Phi_e^+$ and $\Phi_h^+$ that is reflected with unit probability into
a superposition of $\Phi_e^-$ and $\Phi_h^-$.  Orthogonal to these
uncoupled superpositions are the \emph{relevant modes} $\Phi^+_1 = u_e
\Phi_e^+ + u_h \Phi_h^+$ and
$\Phi^-_1 = v_e \Phi_e^- + v_h \Phi_h^-$, which are coupled to the propagating modes in the
superconductor.  We can find them from Eq.~\eqref{eq:smatrix1} by just
observing what $S^\dagger$ and $S$ take $(0,0,1)^\dagger$ to:
\begin{align}
\label{eq:psi1_inout_def}
\begin{pmatrix} u_e  \\ u_h\end{pmatrix}&
= \frac{1}{\mathcal{N}}  
\begin{pmatrix} e^{-i\chi}  \\i {\mathfrak{r}'}^\ast e^{i\chi} \end{pmatrix};&
\begin{pmatrix} v_e  \\ v_h\end{pmatrix}&
= \frac{1}{\mathcal{N}}  
 \begin{pmatrix} i {\mathfrak{r}'}^\ast e^{i\chi'} \\e^{-i\chi'}  \end{pmatrix},
\end{align} 
where $\mathcal{N}=\sqrt{2-\Gamma_1}$ is a normalizing
factor. For our setup, all phase factors here can be absorbed into the definitions of
the plane wave modes in contact $\mathrm{N}_1$, and we obtain Eqs.~\eqref{Phieh}.

Acting with $S$ on $(u_e^\ast,u_h^\ast,0)^\dagger$ allows us to infer the
transmission and reflection amplitudes of the relevant modes, from which we can obtain the
transfer matrix,
\begin{align} 
\label{eq:M_tunnel1}
\begin{pmatrix} \phi_e^+  \\ \phi_h^-\end{pmatrix}&
= \mathcal{M}_1 
\begin{pmatrix} \Phi_1^+  \\ \Phi_1^-\end{pmatrix};&
\mathcal{M}_1 &= 
\frac {1+ (1-\Gamma_1) \sigma_y}{\sqrt{\Gamma_1(2-\Gamma_1)}}.
\end{align}  
This transfer matrix can be written in a succint form 
with a real parameter $\gamma_1$ characterizing the tunnel barrier: 
\begin{align} 
\label{eq:gamma}
\mathcal{M}_1 &= \exp[\gamma_1 \sigma_y]; \quad
\gamma_1 = \frac{1}{2}\ln \frac{2-\Gamma_1}{\Gamma_1}.
\end{align} 
This and the analagous calculation for the right edge of the system
lead directly to Eq.~\eqref{Mgamma}.

\end{document}